\documentclass[
]{ceurart}

\usepackage{listings}
\usepackage{paralist}
\usepackage{enumitem}
\setlist[itemize]{noitemsep}
\setlist[enumerate]{noitemsep}
\setlist[enumerate]{topsep=2pt, itemsep=1pt, parsep=0pt}
\usepackage{cleveref}
\usepackage{wrapfig}
\usepackage{booktabs}
\usepackage{caption}
\usepackage[table]{xcolor}
\usepackage{amssymb} % For \uparrow and \downarrow

\usepackage[most]{tcolorbox}
\tcbset{
  colback=gray!5!white, 
  colframe=gray!50!black, 
  colbacktitle=black,
  coltitle=white,
  fonttitle=\bfseries\footnotesize,
  boxrule=0.5pt, 
  arc=3pt, 
  left=6pt, 
  right=6pt, 
  top=2pt, 
  bottom=2pt, 
  boxsep=4pt,
  enhanced,
  before upper={\vspace{-3pt}},
  after upper={\vspace{-3pt}},
}
\begin{document}

\copyrightyear{2025}
\copyrightclause{Copyright for this paper by its authors. Use permitted under Creative Commons License Attribution 4.0 International (CC BY 4.0).}
\conference{ISWC 2025 Companion Volume, November 2--6, 2025, Nara, Japan}

\conference{ISWC'25: International Semantic Web Conference, Nara, Japan}

\title{Avoiding Over-Personalization with Rule-Guided Knowledge Graph Adaptation for LLM Recommendations}

\author[1]{Fernando Spadea}[%
orcid=0009-0006-4278-3666,
email=spadef@rpi.edu,
]
\cormark[1]
\address[1]{Rensselaer Polytechnic Institute, Troy, NY, USA}

\author[2]{Oshani Seneviratne}[%
orcid=0000-0001-8518-917X,
email=senevo@rpi.edu,
% url=https://oshani.info,
]

\cortext[1]{Corresponding author.}

\begin{abstract}
We present a lightweight neuro-symbolic framework to mitigate over-personalization in LLM-based recommender systems by adapting user-side Knowledge Graphs (KGs) at inference time. Instead of retraining models or relying on opaque heuristics, our method restructures a user’s Personalized Knowledge Graph (PKG) to suppress feature co-occurrence patterns that reinforce Personalized Information Environments (PIEs), i.e., algorithmically induced filter bubbles that constrain content diversity. These adapted PKGs are used to construct structured prompts that steer the language model toward more diverse, \emph{Out-PIE recommendations} while preserving topical relevance. 
We introduce a family of symbolic adaptation strategies, including soft reweighting, hard inversion, and targeted removal of biased triples, and a client-side learning algorithm that optimizes their application per user. Experiments on a recipe recommendation benchmark show that personalized PKG adaptations significantly increase content novelty while maintaining recommendation quality, outperforming global adaptation and naive prompt-based methods.
\end{abstract}

\begin{keywords}
Personalized Knowledge Graphs\sep
Large Language Models \sep
Recommendation Systems\sep
Over-Personalization\sep
Filter Bubbles\sep
Neuro-Symbolic AI\sep
Knowledge Graph Adaptation\sep
User-Centric AI\sep
Recommendation Diversity
\end{keywords}

\maketitle

\section{Introduction}

Overly tailored recommendations lead to Personalized Information Environments (PIEs), which are algorithmically reinforced content silos, akin to filter bubbles, where new information is repeatedly filtered through prior user preferences~\citep{xi2025bursting}. Although PIEs can initially enhance relevance, they often narrow exposure to diverse content, reduce user agency, and inhibit discovery~\citep{pariser2011filter,burrell2016machine}.  Without transparency or control mechanisms, users struggle to diversify their content landscape~\citep{nguyen2014exploring,milano2020recommender,zhang2020explainable}.
Many early approaches to filter bubble mitigation rely on KG embeddings~\citep{donkers2021dual} or influence-based retraining~\citep{anand2022mitigating}, which can be effective but often require retraining, complex domain modeling, or hardcoded logic. 

Building on our prior work~\cite{spadea2025bursting}, we introduce a lightweight inference-time method for escaping PIEs without retraining the underlying language model. By adapting a user's Personalized Knowledge Graph (PKG), which is a structured, editable representation of preferences, we enable interpretable, symbolic control over the recommendation process. Our rule-based PKG adaptation selectively down-weights overrepresented features in the user’s profile, steering large language models (LLMs) to generate more novel recommendations while preserving relevance.

\begin{figure*}[t]
    \centering
    \includegraphics[width=\textwidth]{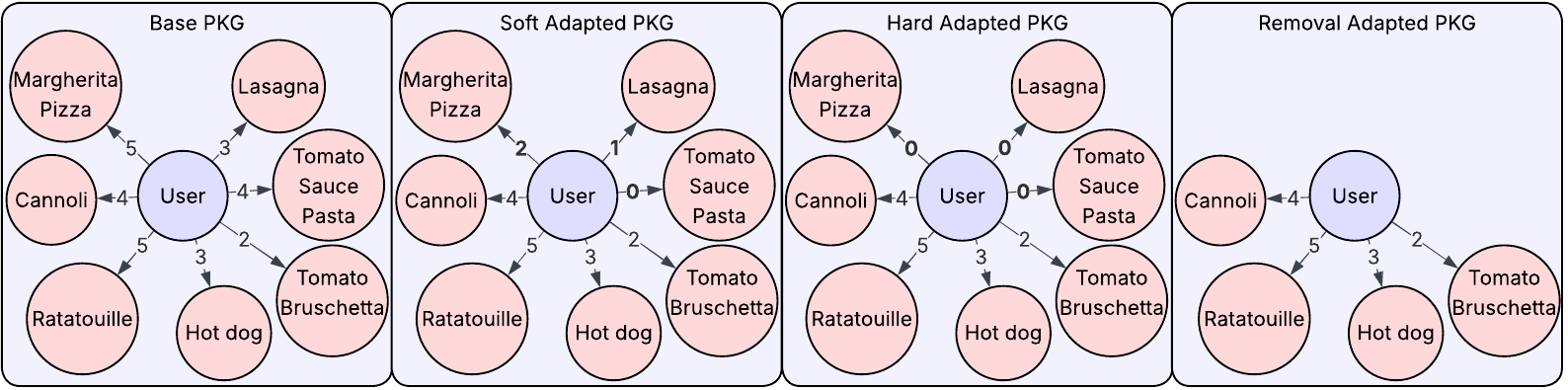}
    \caption{
    Motivating example. The \emph{Base PKG} shows a strong association between Italian cuisine and tomato-based dishes, forming a PIE (filter bubble). The three adaptation strategies modify this bias differently: \emph{Soft} reduces ratings while preserving order, \emph{Hard} inverts ratings more aggressively, and \emph{Removal} deletes PIE-aligned items, such as ``Margherita Pizza,'' without modifying unrelated or already disliked items.}
    % Motivating example. The \emph{Base PKG} illustrates a user's strong positive association between Italian cuisine and tomato-based recipes, creating a PIE, i.e., a filter bubble, around tomato-containing dishes.
    % The \emph{Soft Adapted PKG} reduces preference scores for tomato-based Italian dishes while preserving relative preference ordering, gently steering recommendations toward Italian dishes without tomato.
    % The \emph{Hard Adapted PKG} aggressively replaces the ratings of tomato-based Italian dishes with extremely low scores, strongly discouraging tomato-based Italian dish recommendations.
    % The \emph{Removal Adapted PKG} eliminates highly rated tomato-based Italian dishes from the PKG, such as ``Tomato Pasta'' and ``Margherita Pizza,'' while retaining items like ``Ratatouille'' (non-Italian) and ``Tomato Bruschetta'' (already rated low), thereby narrowing the user's profile to discourage PIE-aligned recommendations without discarding neutral or unrelated content.}
    \label{fig:adapt}
\end{figure*}

\paragraph{Motivating Example:}
Consider a user whose preferences are captured by a PKG. Over time, their interactions with a recommender system create biased associations, forming a PIE. For instance, as shown in \Cref{fig:adapt} \emph{Base PKG}, a user may consistently give high ratings to Italian dishes containing tomatoes, such as ``Tomato Pasta,'' ``Lasagna,'' and ``Margherita Pizza.'' Consequently, the user's PKG becomes dominated by a strong association between ``Italian'' cuisine and the feature ``Tomato.'' Although initially beneficial for relevance, this over-personalization ultimately limits novelty. Ideally, when seeking new Italian recipes, the user would appreciate recommendations that introduce variety, such as ``Pesto Pasta,'' which align with their preference for Italian dishes but break the entrenched link with tomato-based recipes.

\section{Methodology}
\label{sec:method}

\begin{wrapfigure}{r}{0.6\textwidth}
    \centering
    \vspace{-4em}
    \includegraphics[width=0.6\textwidth]{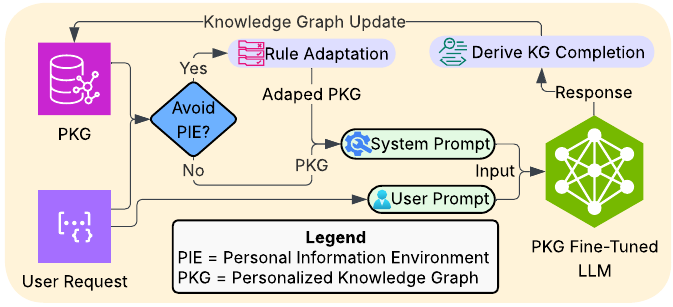}
    \caption{System overview.}
    \label{fig:PKG}
    \vspace{-1em}
\end{wrapfigure}

\Cref{fig:PKG} provides an overview of our recommendation pipeline. When a recommendation query is issued (e.g., ``Suggest an Italian dish''), the system first checks whether the query risks reinforcing a known over-personalization, or PIE. If so, a soft/hard/removal adaptation is applied to adapt the PKG, selectively modifying it to reduce the influence of overrepresented feature pairs (such as ``Italian + tomato'').
The adapted PKG is used to construct a structured prompt that guides the LLM. This prompt reflects the user’s preferences but intentionally avoids PIE-aligned content. The LLM then generates novel, relevant recommendations that satisfy the user's query.

\subsection{Prompt Construction}

The prompt consists of two parts: a system message and a user message. The system message instructs the model to perform KG completion and provides the PKG. It is formatted as follows:

\begin{tcolorbox}[title=System Message, coltitle=white, fonttitle=\bfseries, sharp corners=southwest, enhanced]
\ttfamily\small
You perform Knowledge Graph Completion. You will recommend a new triple to add to the user's knowledge graph with a tail entity that isn't already in their knowledge graph. The user's entity is represented by \{User ID\}. Use this knowledge graph when responding to their queries: \{Knowledge Graph\}
\end{tcolorbox}

The user message requests a recommendation with a specific trait. For our train and test dataset, it is formatted as:

\begin{tcolorbox}[title=User Message, coltitle=white, fonttitle=\bfseries, sharp corners=southwest, enhanced]
\ttfamily\small Recommend a recipe with trait of \{Relation Type\} -> \{Trait Value\}.
\end{tcolorbox}

In the user message, \texttt{Relation Type} is either \texttt{hasIngredient} (for ingredients) or \texttt{hasTag} (for Food.com tags), and \texttt{Trait Value} corresponds to a specific ingredient or tag.

\subsection{Detecting PIEs}

Our system addresses over-personalization by adapting the user's PKG that contains preferences derived from prior interactions (e.g., ratings of recipes annotated with ingredients or tags).

We define a PIE as a user-specific bias toward certain co-occurring pairs of features. Given a pair $(F_{\text{given}}, F_{\text{bias}})$, we detect a PIE if a user consistently assigns significantly higher or lower ratings to items containing both features compared to items containing only $F_{\text{given}}$. For example, as illustrated by the \emph{Base PKG} in \Cref{fig:adapt}, the pair $(F_{\text{given}} = \texttt{Italian}, F_{\text{bias}} = \texttt{tomato})$ forms a positively biased feature association, creating a ``tomato-centric'' PIE. Our objective is to detect these biases and adapt the PKG accordingly to generate recommendations that remain relevant but introduce greater diversity. We quantify the strength of a PIE using a feature-pair bias score, $q_{\text{bias}}$:

\begin{equation}
q_{\text{bias}}(F_{\text{given}}, F_{\text{bias}}) = 
\frac{1}{\mu_{\text{neutral}} \cdot |\mathcal{O}_{F_{\text{given}}}|}
\sum_{o \in \mathcal{O}_{F_{\text{given}}, F_{\text{bias}}}} \left(R_u(o) - \mu_{\text{neutral}}\right)
\label{eq:feature-bias-score}
\end{equation}

Here, $\mu_{\text{neutral}}$ serves as a baseline for interpreting user preferences (e.g., 2.5 on a 0--5 scale), and $R_u(o)$ is the rating that user $u$ assigns to item $o$. The set $\mathcal{O}_{F_{\text{given}}}$ includes all items in the user's PKG that contain the feature $F_{\text{given}}$, while $\mathcal{O}_{F_{\text{given}}, F_{\text{bias}}}$ further restricts this to items that also include $F_{\text{bias}}$.

The bias score $q_{\text{bias}}$ captures how much the user's ratings for items with both features deviate from the neutral point, relative to the size of the user's PKG. A high positive score indicates user preference amplification for feature pairs (e.g., Italian dishes with tomato), while a strong negative score indicates underrating. When $q_{\text{bias}}$ exceeds a set threshold (e.g., $\pm 0.5$), we consider the feature pair to be PIE-inducing and subject to adaptation.

\subsection{Adapting PKGs}
\label{sec:adapting-pkg}

Upon detecting a PIE, we apply one of three \textbf{symbolic adaptation strategies} to selectively modify the PKG before passing it to the LLM, as illustrated in \Cref{fig:adapt}:
\begin{itemize}
    \item \textbf{Soft Adaptation:} Adjusts ratings of PIE-aligned items by symmetrically inverting their strength around a neutral midpoint, preserving relative preference order. For instance, highly rated tomato-based Italian dishes like ``Margherita Pizza'' (rating 5) become somewhat lightly rated (rating 2), while ``Tomato Sauce Pasta'' (rating 4) gets a harsher rating (rating 1), gently nudging recommendations away from tomato dishes while maintaining the user's cuisine preference.
    
    \item \textbf{Hard Adaptation:} Aggressively assigns the extreme opposite ratings to PIE-aligned items. For example, dishes previously highly favored by the user, such as ``Margherita Pizza'' (rating 5) and ``Tomato Sauce Pasta'' (rating 4), receive the lowest possible rating (rating 0), strongly discouraging their recommendation.
    
    \item \textbf{Removal Adaptation:} Completely eliminates PIE-aligned triples from the PKG. For example, recipes like ``Margherita Pizza,'' which explicitly links Italian cuisine and tomato, are entirely removed, forcing the recommender system to explore alternative items.
\end{itemize}

\paragraph{PIE Characterization:}

\begin{itemize}
    
    \item \textbf{Out-PIE:} contains $F_{\text{given}}$ but not $F_{\text{bias}}$; relevant to the user's stated interest but breaks the learned over-personalization.
    \emph{Example:} Cannoli (Italian, no tomato) \textbf{[preferred outcome]}

    \item \textbf{In-PIE:} contains both $F_{\text{given}}$ and $F_{\text{bias}}$; reinforces the over-personalized association the user is trying to avoid.
    \emph{Example:} Margherita pizza (Italian, tomato)
    
    \item \textbf{Invalid:} does not contain $F_{\text{given}}$; fails to satisfy the user’s original intent or query (e.g., recommending a non-Italian dish when the user asked for Italian).
    \emph{Example:} Ratatouille (French, tomato)
\end{itemize}

\paragraph{Tuning the PKG Adaptation Proportion:}

To control the extent of symbolic intervention, we introduce a user-specific parameter called \texttt{adaptProportion}, which determines the fraction of PIE-aligned triples in the PKG that should be adapted before inference. A low \texttt{adaptProportion} results in minimal intervention, while a high \texttt{adaptProportion} aggressively steers the recommendation away from the PIE.
 
We learn a personalized \texttt{adaptProportion} for each user using a feedback-driven tuning algorithm, which simulates PIE-avoidance scenarios using synthetic data points, and incrementally adjusts \texttt{adaptProportion} based on their outcomes. If the adapted PKG still yields \texttt{In-PIE} recommendations, the proportion is increased. If it produces \texttt{Invalid} results, the proportion is decreased. Successful \texttt{Out-PIE} results leave the parameter unchanged. This iterative procedure converges on a personalized adaptation strength tailored to each user's PKG structure and feature biases.

\subsection{Model Fine-Tuning}

We fine-tune Qwen3-0.6B~\cite{huggingface2024qwen3} using Hugging Face’s \texttt{KTOTrainer}~\cite{huggingface2024kto}, which implements Kahneman-Tversky Optimization (KTO)~\cite{ethayarajh2024kto}.
Each training data point consists of a \textit{(prompt, completion, label)} triplet: the prompt includes user context and a query, the completion is a potential response to the prompt, and the label is a binary indicator of the completion's quality (positive or negative). 

Training data is derived from a customized version of the \textit{Food.com Recipes and Interactions dataset}~\cite{shuyang_li_2019}, which includes user ratings (on a 0–5 star scale), ingredients, and categorical tags for recipes. From this corpus, we construct the PKG capturing individual user preferences as a set of rated recipes.

\section{Experimental Results}

\begin{wraptable}{r}{0.6\textwidth}
    \vspace{-4em}
    \centering
    \caption{Performance of PKG adaptation strategies. 
    {\color{green!50!black}↑} better for \textbf{Out-PIE}; 
    {\color{green!60!black}↓} better for \textbf{In-PIE}, and \textbf{Invalid}. The best values are highlighted in \colorbox{green!20}{green}, and the worst values are highlighted in \colorbox{red!20}{red}.}
    \label{tab:cen}
    \setlength{\tabcolsep}{3pt}
    \begin{tabular}{lccc}
        \toprule
        \textbf{Strategy} & \textbf{Out-PIE} {\color{green!50!black}↑} & \textbf{In-PIE} {\color{green!60!black}↓} & \textbf{Invalid} {\color{green!60!black}↓} \\
        \midrule
        Soft (personalized)       & \cellcolor{green!8}0.3237 & \cellcolor{green!5}0.2158 & \cellcolor{green!10}0.4604 \\
        Soft (global)             & 0.2517 & \cellcolor{red!10}0.2583 & 0.4901 \\
        \midrule
        Hard (personalized)       & 0.2848 & 0.2152 & 0.5000 \\
        Hard (global)             & 0.2768 & \cellcolor{green!5}0.1977 & 0.5254 \\
        \midrule
        Removal (personalized)    & 0.3020 & 0.2416 & \cellcolor{green!5}0.4564 \\
        Removal (global)          & \cellcolor{green!10}0.3277 & \cellcolor{green!5}0.2203 & \cellcolor{green!5}0.4520 \\
        \midrule
        Prompt-Based Adaptation             & \cellcolor{red!15}0.1925 & \cellcolor{green!10}0.1863 & \cellcolor{red!20}0.6211 \\
        \midrule
        No Adaptation                      & 0.2517 & \cellcolor{red!10}0.2583 & 0.4901 \\
        \bottomrule
    \end{tabular}
    \vspace{-1em}
\end{wraptable}

We evaluate our PIE avoidance framework using PKGs derived from the \textit{Food.com Recipes and Interactions dataset}~\cite{shuyang_li_2019}. We randomly select 20 user PKGs and, for each, sample 50 PIE-inducing feature pairs. These are split 80/20 into 40 training and 10 evaluation PIEs per user. The training set is used to learn a personalized \texttt{adaptProportion} for each user via the tuning algorithm in \Cref{sec:adapting-pkg}. For comparison, we also compute a single global \texttt{adaptProportion} across all 800 training PIEs. A learning rate of 0.05 is used in both cases.

During evaluation, we generate 10 test queries per user (200 total) and categorize each model-generated recommendation into \texttt{Out-PIE}, \texttt{In-PIE}, or \texttt{Invalid} classes. We aggregate and normalize the counts to obtain proportions summing to one.

We benchmark three PKG adaptation strategies (soft, hard, removal), each with both personalized and global \texttt{adaptProportion} settings, against two baselines:
(1) a prompt-based method using natural language instructions to avoid PIEs, and
(2) a no-adaptation baseline.

As shown in \Cref{tab:cen}, soft adaptation with personalized tuning achieves the best overall performance. It increases \texttt{Out-PIE} recommendations from 25.2\% (global \texttt{adaptProportion} baseline) to 32.4\%, while simultaneously reducing the \texttt{Invalid} rate from 49.0\% to 46.0\%, avoiding over-personalization without decreasing recommendation quality. 
In contrast, the prompt-based approach, which attempts PIE avoidance via plain-text instructions, performs poorly, with the lowest \texttt{Out-PIE} rate (19.3\%) and the highest \texttt{Invalid} rate (62.1\%). This underscores the limitations of relying solely on natural language prompting and demonstrates the effectiveness of symbolic PKG adaptations.

\section{Conclusion}

We introduced a novel, neuro-symbolic framework for enhancing personalization and user agency in recommender systems by adapting PKGs rather than modifying model internals. Unlike conventional approaches that require model retraining or rely on brittle heuristics, our method operates entirely at the user-side knowledge representation layer, enabling efficient, interpretable, and privacy-preserving control over recommendation behavior.

Our key contributions include: (1) a formalization of PIEs as measurable feature-pair biases within PKGs, (2) a suite of symbolic PKG adaptation strategies (Soft, Hard, and Removal) that steer LLMs toward more diverse, Out-PIE content, and (3) a client-side learning algorithm for optimizing user-specific adaptation policies. Through an evaluation on a real-world recipe dataset, we show that personalized PKG adaptation consistently outperforms both global adaptation and natural-language prompting in reducing over-personalization without compromising recommendation quality.

These findings point toward a broader paradigm shift: from tuning black-box models to shaping the symbolic structures that guide them. Our work demonstrates that adapting structured user knowledge offers a powerful, generalizable mechanism for embedding user intent and improving controllability in LLM-powered systems, laying the groundwork for safer, more personalized, and user-aligned AI for diverse recommendations.

%% The declaration on generative AI comes in effect
%% https://ceur-ws.org/GenAI/Policy.html
\section*{Declaration on Generative AI}
 During the preparation of this work, the authors used ChatGPT, Gemini and Grammarly in order to rephrase some of the sentences and also to fix grammar and spelling issues. After using these tools and services, the authors reviewed and edited the content as needed and take full responsibility for the publication’s content. 

\section*{Supplemental Materials}\label{apd:first}

All research artifacts, including source code, dataset construction scripts, and result generation pipelines, are available in our GitHub repository. All external datasets and software dependencies used in this work are documented and linked in the repository’s README.\\
\url{https://github.com/brains-group/KGAdaptation}. 

%%
%% Define the bibliography file to be used
\bibliography{references}

\end{document}